\documentclass[twocolumn,floatfix,aps]{revtex4-2}
\usepackage{graphicx}
\usepackage{amsmath}
\usepackage{amssymb}
\usepackage{bm}
\usepackage{hyperref}

\usepackage{color}

\begin{document}

\title{Repulsive Casimir force from a Majorana zero-mode}
\author{C. W. J. Beenakker}
\affiliation{Instituut-Lorentz, Universiteit Leiden, P.O. Box 9506, 2300 RA Leiden, The Netherlands}
\date{February 2024}

\begin{abstract}
Fu and Kane have taught us that a Majorana zero-mode appears on the quantum spin Hall edge at the interface with a superconductor. If a magnetic scatterer is placed on the edge, the zero-point energy of massless edge excitations exerts a force on the scatterer. This is the fermionic analogue of the electromagnetic Casimir effect. We show that the Majorana zero-mode produces a repulsive Casimir force, pushing the scatterer away from the superconductor. Unlike some other signatures of Majorana zero-modes, the repulsive Casimir force is directly tied to the topological invariant of the system (the sign of the determinant of the reflection matrix from the superconductor).
\end{abstract}
\maketitle

\section{Introduction}
A Majorana zero-mode is a topological defect in a superconductor \cite{Kit01}. It is described formally as a state that is annihilated by the Hamiltonian, or informally as ``half an electron'' \cite{Lei12,Ali12,Bee13}. It may find applications in quantum information processing \cite{Nay08,Das15}, but it is elusive \cite{Fle21,Das23}: Although the zero-mode is a mid-gap state that can be detected spectroscopically \cite{Pra20}, there are confounding factors in a superconductor that may produce a non-topological mid-gap spectral peak \cite{Che19}. 

One reason why probes of Majorana zero-modes can be ambiguous is that they do not directly connect to a topological invariant. In a scattering formulation the topological invariant of a superconductor is the sign of the determinant of the reflection matrix \cite{Akh11}. A negative determinant identifies an unpaired Majorana zero-mode. Here we wish to show how the fermionic Casimir effect can provide such a probe.

The electromagnetic Casimir effect \cite{Cas48,Plu86,Mos97} refers to the attractive force between two metal plates produced by the zero-point energy of photons. In the fermionic analogue \cite{Sun04} the photons are replaced by massless excitations with the same linear dispersion relation, as they appear in graphene or in topological insulators \cite{Zha08,Shy09,Woo16,Lu21,Ish21,Bee24}. Non-topological Majorana fermions (not the zero-modes) have been studied in the context of neutrino physics \cite{Oik10,Che10} --- the Casimir effect has been proposed as a way to distinguish Dirac fermions from Majorana fermions \cite{Cos20}.

\begin{figure}[tb]
\centerline{\includegraphics[width=0.9\linewidth]{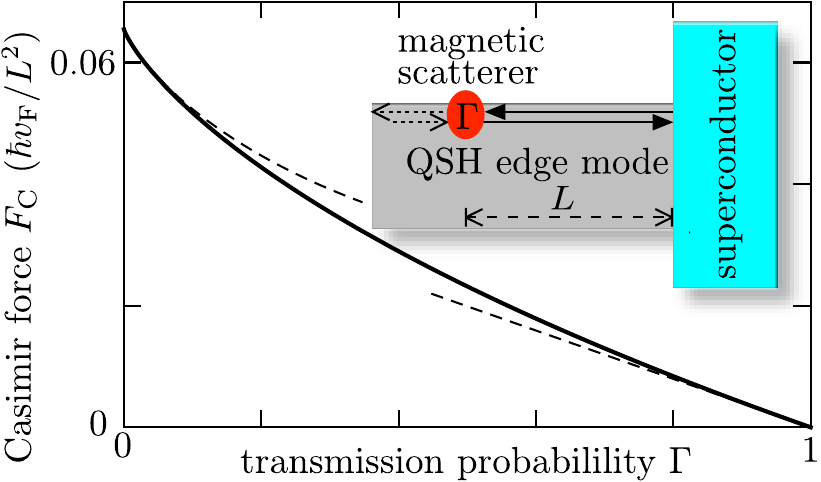}}
\caption{Casimir force on a magnetic scatterer, placed on the edge of a quantum spin Hall (QSH) insulator at a distance $L$ from a superconductor. The force points away from the superconductor. The plot gives the dependence of the Casimir force on the transmission probability $\Gamma$ of the QSH edge mode, in the long-distance regime $L\gg\xi_0$ at zero temperature (Eq.\ \eqref{FCasymptotes}, with the asymptotes dashed).
}
\label{fig_gamma}
\end{figure}

To investigate how the Casimir effect is modified by the presence of a topological Majorana zero-mode we consider the geometry of Fig.\ \ref{fig_gamma} (inset): A segment of length $L$ of the conducting edge of a quantum spin Hall insulator is gapped at one end by a superconductor (\textit{s}-wave gap $\Delta_0$).  At the other end it is gapped by a magnetic scatterer, which could be the tip of a scanning probe. As pointed out by Fu and Kane \cite{Fu09}, this topologically nontrivial system hosts a pair of Majorana zero-modes, one on each edge. The two edges are decoupled by the insulating bulk, the magnetic scatterer is only influenced by the Majorana zero-mode on the upper edge.  

The magnetic scatterer will experience a Casimir force, given by the derivative $F_{\rm C}=-d{\cal F}/dL$ of the free energy with respect to the separation $L$ to the superconductor. The force can be attractive, pointing towards the superconductor, or repulsive, pointing away from it. We will show that a repulsive Casimir force is a signature of topological superconductivity and the presence of a Majorana zero-mode.

\section{Scattering formulation}
The free energy in equilibrium at temperature $T$ is given by
\begin{equation}
{\cal F}=-T\int_0^\infty d\varepsilon\,\rho(\varepsilon)\ln\bigl[2\cosh(\varepsilon/2T)\bigr],\label{calFdef}
\end{equation}
with $\rho(\varepsilon)$ the density of states of Bogoliubov quasiparticle excitations (electrons and holes) at energy $\varepsilon$, relative to the Fermi level. Spin and valley degeneracies can be included in the density of states, but on the quantum spin Hall edge there are no degeneracy factors. We set $\hbar$ and Boltzmann's constant $k_{\rm B}$ equal to unity in most equations.

In a scattering formulation \cite{Bro97} the integral over real energies can be transformed into a sum over imaginary Matsubara frequencies, see App.\ \ref{app_free_energy}:
\begin{equation}
\begin{split}
&{\cal F}=-T\sum_{p=0}^\infty\ln\det[1-R_{\text{M}}(i\omega_p)R_{\text{S}}(i\omega_p)]+{\cal F}_{\rm free},\\
&\omega_p=(2p+1)\pi T.
\end{split}
\label{calFgeneral}
\end{equation}
The additional term ${\cal F}_{\rm free}$ is the free energy of the unconfined quantum spin Hall edge, which does not contribute to the Casimir force and will be omitted in what follows. 

The change in the free energy due to the confinement is expressed in terms of the product of reflection matrices $R_{\text{M}}(\varepsilon)$ and $R_{\text{S}}(\varepsilon)$ from the magnetic scatterer (M) and the superconductor (S).  Electron-hole symmetry requires that
\begin{equation}
R_{\text{M,S}}(-\varepsilon)=\nu_x R_{\text{M,S}}^\ast(\varepsilon)\nu_x,\label{Rsymm}
\end{equation}
with $\nu_x$ a Pauli matrix in the electron-hole degree of freedom.

Time-reversal symmetry forbids back scattering along the gapless quantum spin Hall edge \cite{Kan05}. Transmission over a distance $L$ introduces a phase shift $k(\varepsilon)L$ for the electron and $-k(-\varepsilon)L$ for the hole. The resulting $L$-dependence of the free energy is introduced via the substitution
\begin{equation}
R_{\text{M}}R_{\text{S}}=U\tilde{R}_{\text{M}}U\tilde{R}_{\text{S}},\;\;U=\begin{pmatrix}
e^{ik(\varepsilon)L}&0\\
0&e^{-ik(-\varepsilon)}
\end{pmatrix}.
\end{equation}
The reflection matrices $\tilde{R}_{\text{M}}$ and $\tilde{R}_{\text{S}}$ are $L$-independent.

Near the Fermi energy $E_{\rm F}$ we may linearize $k(\varepsilon)=k_{\rm F}+\varepsilon/v_{\rm F}$, hence
\begin{equation}
U(\varepsilon)=e^{i\varepsilon L/v_{\rm F}}e^{i\nu_z k_{\rm F}L}.
\end{equation}
Taking also the zero-temperature limit the free energy becomes
\begin{equation}
{\cal F}=-\int_0^\infty \frac{d\omega}{2\pi}\,\ln\det[1-e^{-2\omega L/v_{\rm F}}e^{2i\nu_z k_{\rm F}L}\tilde{R}_{\text{M}}(i\omega)\tilde{R}_{\text{S}}(i\omega)],\label{calFzeroT1}
\end{equation}
where we have used that $\tilde{R}_{\rm M}$ commutes with $\nu_z$ (because the magnetic scatterer does not couple electrons and holes).

\section{Casimir force}
The Casimir force depends on the ratio $L/\xi_0$ of the junction length and the superconducting coherence length $\xi_0=\hbar v_{\rm F}/\Delta_0$. In the long-distance regime $L\gg\xi_0$ one may neglect the energy dependence of the reflection matrices, evaluating them at $\varepsilon=0=\omega$,
\begin{equation}
\begin{split}
&{\cal F}=-\int_0^\infty \frac{d\omega}{2\pi}\,\ln\det(1-e^{-2\omega L/v_{\rm F}}\Omega),\\
&\Omega=e^{2i\nu_z k_{\rm F}L}\tilde{R}_{\text{M}}(0)\tilde{R}_{\text{S}}(0).
\end{split}\label{calFOmega}
\end{equation}

We assume that $\Omega$ is unitary at $\varepsilon=0$ (no transmission through the magnetic scatterer, we will relax this assumption later on). The electron-hole symmetry relation 
\begin{equation}
\Omega^\ast=\nu_x\Omega\nu_x
\end{equation}
then implies that 1) if $\lambda$ is an eigenvalue of $\Omega$, then also $\lambda^\ast=1/\lambda$ is an eigenvalue; and 2) the determinant of $\Omega$ equals $\pm 1$. The determinant is $+1$ if the superconductor is topologically trivial, while $\det\Omega=-1$ if the superconductor is topologically nontrivial \cite{Akh11}. The sign change is due to reflection from an unpaired Majorana zero-mode. 

On the quantum spin Hall edge, the condition $\det\Omega =-1$ pins the two eigenvalues at $+1$ and $-1$, independent of $L$. The free energy then evaluates to
\begin{align}
{\cal F}={}&-\frac{1}{2\pi}\int_0^\infty d\omega\,\sum_{s=\pm 1}\ln\bigl(1+se^{-2\omega L/v_{\rm F}}\bigr)\nonumber\\
={}&-\frac{1}{2\pi}\int_0^\infty d\omega\,\ln\bigl(1-e^{-4\omega L/v_{\rm F}}\bigr)=\frac{\pi\hbar v_{\rm F}}{48 L}.\label{F48}
\end{align}
This corresponds to a repulsive Casimir force $F_{\rm C}\propto L^{-2}$ on the scatterer.

\section{Trivial versus nontrivial superconductor}
To contrast this with a conventional, trivial superconductor, we consider a more general case in which the gapless region of length $L$ may have multiple conducting modes $n=1,2,\ldots N$ (counting degeneracies). We still assume that there is no backscattering upon propagation over the length $L$, which in the general case without topological protection will require $L$ small compared to the mean free path. 

The wave vector $k_n$ will in general be mode dependent, but the dependence is weak for the modes with the largest longitudinal velocity (which give the dominant contribution to ${\cal F}$). For simplicity we neglect the $n$-dependence of $k_n$. We thus arrive at the same equation \eqref{calFOmega}, but now $\Omega$ is a $2N\times 2N$ unitary matrix. We consider the free energy ${\cal F}_\pm$ separately in the two cases $\det\Omega=\pm 1$.

For a trivial superconductor, if $\det \Omega=+1$, the $2N$ eigenvalues of $\Omega$ come in $N$ complex conjugate pairs $e^{\pm i\phi_n}$, with $0\leq \phi_n\leq \pi$. The free energy then evaluates to
\begin{align}
{\cal F}_+={}&-\frac{1}{\pi}\sum_{n=1}^N\operatorname{Re}\int_0^\infty d\omega\,\ln\bigl(1+e^{-2\omega L/v_{\rm F}}e^{i\phi_n}\bigr)\nonumber\\
={}&-\frac{\pi\hbar v_{\rm F}}{24 L}\sum_{n=1}^N(1-3\phi_n^2/\pi^2).
\end{align}
For a nontrivial superconductor, if $\det \Omega=+1$, two of the eigenvalues of $\Omega$ are pinned on the real axis, one at $+1$ and the other at $-1$. The other eigenvalues are still complex conjugate pairs, hence
\begin{align}
{\cal F}_-={}&\frac{\pi\hbar v_{\rm F}}{48 L}-\frac{\pi\hbar v_{\rm F}}{24 L}\sum_{n=2}^N(1-3\phi_n^2/\pi^2).
\end{align}

If the superconductor is disordered the scattering phases $\phi_n$ will be uniformly distributed in $(0,\pi)$, and we may average
\begin{equation}
\langle 1-3\phi^2/\pi^2\rangle=\int_0^\pi(1-3\phi^2/\pi^2)\frac{d\phi}{\pi}=0
\end{equation}
Then $\langle {\cal F}_+\rangle=0$ while $\langle {\cal F}_-\rangle= \pi\hbar v_{\rm F}/48 L$. All of this assumes that backscattering upon propagation between the barriers can be neglected.

\section{Model calculation}
The precise formula \eqref{F48} for the repulsive Casimir force from a Majorana zero-mode assumes limiting conditions (long distance, fully reflecting magnetic scatterer, zero temperature). We may relax these conditions in a model calculation. 

We choose a gauge where the pair potential in the superconductor is real. Its reflection matrix is \cite{Bee13b}
\begin{equation}
\tilde{R}_{\rm S}(\varepsilon)=\alpha(\varepsilon)\nu_x,\;\;\alpha(\varepsilon)=i\varepsilon/\Delta_0+\sqrt{1-\varepsilon^2/\Delta_0^2},\label{RS}
\end{equation}
evaluated at $\varepsilon+i0^+$ to avoid the branch cut of the square root. (Check that the symmetry \eqref{Rsymm} is satisfied.) The off-diagonal Pauli matrix $\nu_x$ signifies Andreev reflection, from electron to hole. Normal reflection (from electron to electron) is forbidden by time-reversal symmetry on the quantum spin Hall edge. The reflection matrix is unitary below the gap, while above the gap there is also propagation into the superconductor and $\tilde{R}_{\rm S}$ decays as $1/\varepsilon$.

The magnetic scatterer has only normal reflection, with probability $1-\Gamma$ and phase shift $\pm\phi$ for electron and hole,
\begin{equation}
\tilde{R}_{\rm M}=e^{i\phi\nu_z}\sqrt{1-\Gamma}.
\end{equation}
For simplicity we do not include the energy dependence of $\tilde{R}_{\rm M}$ (assuming that the $\varepsilon$-dependence is on a scale large compared to $\Delta_0$).

At zero temperature the free energy \eqref{calFzeroT1} is then given by
\begin{equation}
{\cal F}=-\int_0^\infty \frac{d\omega}{2\pi}\,\ln\bigl[1-(1-\Gamma)\alpha(i\omega)^2e^{-4\omega L/v_{\rm F}}\bigr].\label{calFbarrier}
\end{equation}
The phase shifts $k_{\rm F}L$ and $\phi$ drop out of the determinant because of the identity $e^{a\nu_z}\nu_x e^{a\nu_z}=\nu_x$ for any $a\in\mathbb{C}$. 

\begin{figure}[tb]
\centerline{\includegraphics[width=0.8\linewidth]{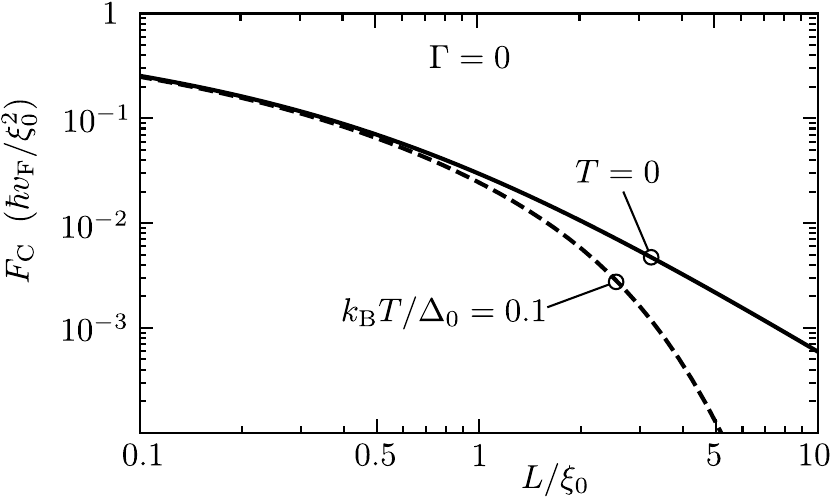}}
\caption{Distance dependence of the Casimir force at zero and at nonzero temperature, on a log-log scale.
}
\label{fig_temperature}
\end{figure}

The resulting Casimir force $F_{\rm{C}}=-d{\cal F}/dL$ has in the long-distance regime the asymptotics
\begin{subequations}
\begin{align}
F_{\rm C}={}&\frac{\hbar v_{\rm F}}{L^2}\frac{\text{Li}_2(1-\Gamma)}{8 \pi },\;\;\text{for}\;\;L\gg\xi_0,\\
={}&\frac{\hbar v_{\rm F}}{8\pi L^2}\times\begin{cases}
(\pi^2/6+\Gamma \ln \Gamma-\Gamma)&\text{if}\;\;\Gamma\ll 1,\\
(1-\Gamma)&\text{if}\;\;\Gamma\lesssim 1.
\end{cases}
\end{align}\label{FCasymptotes}
\end{subequations}
(The function $\text{Li}_2$ is the dilogarithm.) In the opposite short-distance regime the asymptotics is
\begin{equation}
F_{\rm C}=\frac{\Delta_0}{2\pi\xi_0}(1-\Gamma)\ln(\xi_0/L),\;\;\text{for}\;\;L\ll\xi_0.
\end{equation}

The Casimir force is repulsive for all $L$. The $1/L^2$ decay for large distances becomes a logarithmic increase for short distances, saturating at the Fermi wave length, when $F_{\rm C}\simeq (\Delta_0/\xi_0)\ln(E_{\rm F}/\Delta_0$). (A similar logarithmic increase appears in a Josephson junction \cite{Bee23}.)

All of this is at zero temperature. At non-zero temperature we sum over the discrete Matsubara frequencies, $\int_0^\infty d\omega\mapsto (2\pi T)\sum_{\omega_p=(2p+1)\pi T}$. The chararacteristic temperature scale is the smallest of $\hbar v_{\rm F}/k_{\rm B}L$ and $\Delta_0/k_{\rm B}$. As shown in Fig.\ \ref{fig_temperature}, for $L\lesssim\xi_0$ the temperature dependence is insignificant for $T\lesssim 0.1\,\Delta_0/k_{\rm B}$.

\section{Conclusion}

Prospects for direct measurement of the repulsive Majorana-Casimir force are not promising: Even in the short-distance regime a force of order $\Delta_0/\xi_0\lesssim 10^{-15}\,\rm{N}$ is beyond present detection capabilities. Conceptually the effect provides a novel connection between zero-point fluctuations of gapless fermionic excitations and the presence of a Majorana zero-mode. The connection involves a topological invariant, so it directly ties to the defining property of a topological superconductor.

It is instructive to compare with other appearances of a repulsive fermionic Casimir effect: in a carbon nanotube \cite{Zha08} and in a Josephson junction \cite{Bee23,Kri04,Par12}. In these topologically trivial systems the Casimir force crosses over from attractive to repulsive as a parameter is varied (the orientation of the magnetization in the carbon nanotube and the superconducting phase difference in the Josephson junction). In contrast, here the sign of the Casimir force is fixed by a topological invariant.

It is also of interest to compare with the electromagnetic Casimir effect in vacuum, which may become repulsive if the mirrors that confine the photons break time reversal symmetry \cite{Woo16,Lu21}. A connection with topological invariants appears if the mirrors are Chern insulators: the Casimir force is attractive or repulsive depending on whether the Chern numbers in the two mirrors have the same or the opposite sign \cite{Rod14}.

\acknowledgments

I have benefited from discussions with A. R. Akhmerov and T. Vakhtel. This project has received funding from the European Research Council (ERC) under the European Union's Horizon 2020 research and innovation programme.

\appendix

\section{Free energy of a magnet--normal--superconductor junction}
\label{app_free_energy}

\begin{figure}[tb]
\centerline{\includegraphics[width=0.8\linewidth]{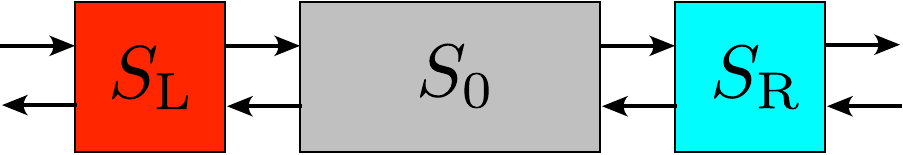}}
\caption{Three scattering regions in series, each with their own unitary scattering matrix.
}
\label{fig_scattering}
\end{figure}

To make the paper self-contained, we give the derivation of the free energy formula \eqref{calFgeneral}. This calculation for a magnet--normal--superconductor junction is analogous to the calculation for a superconductor--normal--superconductor junction of Ref.\ \onlinecite{Bro97}.

\subsection{Scattering formulation}

The scattering geometry is shown schematically in Fig.\ \ref{fig_scattering}. It consists of three scattering regions in series, with scattering matrices $S_{\rm L}$ (tunneling through the magnetic scatterer), $S_0$ (propagation along the edge mode), and $S_{\rm R}$ (Andreev reflection by the superconductor). The scattering regions are connected by leads with $2N$ propagating modes at energy $\varepsilon$. 

In each region we allow for both transmission and reflection, via the block structure
\begin{equation}
S_X=\begin{pmatrix}
r_X&t'_X\\
t_X&r'_X
\end{pmatrix},\;\;X\in\{L,R,0\}.\label{SXdef}
\end{equation}
The blocks contain the reflection submatrices $r_X,r'_X$ (reflection from the left and from the right) and the transmission submatrices $t_X,t'_X$ (transmission from left-to-right and from right-to-left).  These submatrices have dimension $2N\times 2N$. The scattering matrix is a $4N\times 4N$ unitary matrix, $S_X^{\vphantom{\dagger}}S_X^\dagger=1$.

The full scattering matrix ${\cal S}$ of the entire structure is given by
\begin{equation}
{\cal S}={\cal R}+{\cal T}'S_0
(1-{\cal R}'S_0)^{-1}
{\cal T},\label{calSdef}
\end{equation}
where we have defined the block-diagonal matrices
\begin{subequations}
\label{calRTdef}
\begin{align}
&{\cal R}=\begin{pmatrix}
r_{\rm L}&0\\
0&r'_{\rm R}
\end{pmatrix},\;\;{\cal R}'=\begin{pmatrix}
r'_{\rm L}&0\\
0&r_{\rm R}
\end{pmatrix},\\
&{\cal T}=\begin{pmatrix}
t_{\rm L}&0\\
0&t'_{\rm R}
\end{pmatrix},\;\;{\cal T}'=\begin{pmatrix}
t'_{\rm L}&0\\
0&t_{\rm R}\end{pmatrix}.
\end{align}
\end{subequations}

\subsection{Density of states}

Knowledge of the energy dependence of ${\cal S}(\varepsilon)$ gives the density of states via the Friedel formula,
\begin{equation}
\rho(\varepsilon)=\frac{1}{2\pi i}\frac{d}{d\varepsilon}\ln\det {\cal S}(\varepsilon)+\rho_{\rm lead}(\varepsilon),
\end{equation}
where $\rho_{\rm lead}$ is the density of states of the leads connecting the scattering regions. For the Casimir effect we need to subtract the contributions from the individual scattering regions, 
\begin{equation}
\rho_X(\varepsilon)=\frac{1}{2\pi i}\frac{d}{d\varepsilon}\ln\det S_X(\varepsilon).
\end{equation}
For that purpose we note the identities
\begin{equation}
\det S_X=\frac{\det r'_X}{\det r_X^\dagger}=\frac{\det (-t'_X)}{\det t_X^\dagger},\label{detSX}
\end{equation}
which follow directly from the polar decomposition
\begin{equation}
S_X=\begin{pmatrix}
U_X&0\\
0&V_X
\end{pmatrix}\begin{pmatrix}
\sqrt{1-T_X}&-\sqrt{T_X}\\
\sqrt{T_X}&\sqrt{1-T_X}
\end{pmatrix}\begin{pmatrix}
U'_X&0\\
0&V'_X
\end{pmatrix},
\end{equation}
 with unitary matrices $U_X,V_X,U'_X,V'_X$ and a diagonal matrix of transmission probabilities $T_X$.

We now rewrite Eq.\ \eqref{calSdef}, using the unitarity relations
\begin{equation}
{\cal TT}^\dagger+{\cal R}'^{\vphantom{\dagger}}{\cal R}'^\dagger=1,\;\;{\cal R}=-{\cal T}'{\cal R}'^\dagger ({\cal T}^\dagger)^{-1},
\end{equation}
as follows:
\begin{widetext}

\begin{align}
{\cal S}&={\cal R}+{\cal T}'S_0(1-{\cal R}'S_0)^{-1}{\cal T}\nonumber\\
&={\cal T}'[-{\cal R}'^\dagger ({\cal T}^\dagger)^{-1}+S_0(1-{\cal R}'S_0)^{-1}{\cal T}]\nonumber\\
&={\cal T}'[-{\cal R}'^\dagger({\cal TT}^\dagger)^{-1}+S_0(1-{\cal R}'S_0)^{-1}]{\cal T}\nonumber\\
&={\cal T}'[-{\cal R}'^\dagger(1-{\cal R}'^{\vphantom{\dagger}}{\cal R}'^\dagger)^{-1}+S_0(1-{\cal R}'S_0)^{-1}]{\cal T}\nonumber\\
&={\cal T}'[-{\cal R}'^\dagger(1-{\cal R}'^{\vphantom{\dagger}}{\cal R}'^\dagger)^{-1}(1-{\cal R}'S_0)+S_0](1-{\cal R}'S_0)^{-1}{\cal T}\nonumber\\
&={\cal T}'[{\cal R}'^\dagger {\cal R}'^{\vphantom{\dagger}}(1-{\cal R}'^\dagger {\cal R}'^{\vphantom{\dagger}})^{-1}S_0-{\cal R}'^\dagger(1-{\cal R}'^{\vphantom{\dagger}}{\cal R}'^\dagger)^{-1}+S_0](1-{\cal R}'S_0)^{-1}{\cal T}\nonumber\\
&={\cal T}'[(1-{\cal R}'^\dagger {\cal R}'^{\vphantom{\dagger}})^{-1}S_0-(1-{\cal R}'^\dagger {\cal R}'^{\vphantom{\dagger}})^{-1}{\cal R}'^\dagger](1-{\cal R}'S_0)^{-1}{\cal T}\nonumber\\
&={\cal T}'(1-{\cal R}'^\dagger {\cal R}'^{\vphantom{\dagger}})^{-1}(S_0-{\cal R}'^\dagger)(1-{\cal R}'S_0)^{-1}{\cal T}\nonumber\\
&=({\cal T}'^\dagger)^{-1}S_0(1-S_0^\dagger{\cal R}'^\dagger)(1-{\cal R}'S_0)^{-1}{\cal T}.
\end{align}
\end{widetext}
The resulting density of states is
\begin{align}
\rho(\varepsilon)={}&-\frac{1}{\pi}\frac{d}{d\varepsilon}\operatorname{Im}\ln\det (1-{\cal R}'S_0)\nonumber\\
&+\rho_0+\rho_{\rm L}+\rho_{\rm R}+\rho_{\rm lead},\label{rhofull}
\end{align}
where we have used Eq.\ \eqref{detSX} to extract the contributions from the separate scattering regions.

The change in the density of states due to the confinement, $\delta\rho=\rho-\rho_0-\rho_{\rm L}-\rho_{\rm R}-\rho_{\rm lead}$, is then given by the first term in Eq.\ \eqref{rhofull}, which upon substitution of Eqs.\ \eqref{SXdef} and \eqref{calRTdef} takes the form
\begin{equation}
\delta\rho(\varepsilon)=-\frac{1}{\pi}\frac{d}{d\varepsilon}\operatorname{Im}\ln\det\begin{pmatrix}
1-r'_{\rm L}r_0&r'_{\rm L}t'_0\\
r_{\rm R}t_0&1-r_{\rm R}r'_0
\end{pmatrix}.
\end{equation}
In our case the normal region is a quantum spin Hall edge, without backscattering: $r_0=0=r'_0$, hence
\begin{equation}
\begin{split}
&\delta\rho(\varepsilon)=-\frac{1}{\pi}\frac{d}{d\varepsilon}\operatorname{Im}\ln\det(1-R_{\rm M}R_{\rm S}),\\
&R_{\rm M}=r'_{\rm L}t'_0,\;\;R_{\rm S}=r_{\rm R}t_0.\label{deltarhoresult}
\end{split}
\end{equation}

\subsection{Free energy}

The free energy follows upon substitution of Eq.\ \eqref{deltarhoresult} into Eq.\ \eqref{calFdef},
\begin{align}
{\cal F}={}&\frac{T}{\pi}\int_0^\infty d\varepsilon \,\ln\bigl[2\cosh(\varepsilon/2T)\bigr]\nonumber\\
&\times \frac{d}{d\varepsilon}\operatorname{Im}\ln\det[1-R_{\rm M}(\varepsilon)R_{\rm S}(\varepsilon)] +{\cal F}_{\rm free},
\end{align}
where ${\cal F}_{\rm free}$ is the contribution from the separate scattering regions. Because of the particle-hole symmetry relation \eqref{Rsymm} the extension of the integral to $\int_{-\infty}^\infty$ gives $2i$ times the imaginary part of the integral $\int_0^\infty$. Upon partial integration we then have
\begin{equation}
{\cal F}=\frac{i}{4\pi}\int_{-\infty}^\infty d\varepsilon \,\tanh(\varepsilon/2T)\ln\det[1-R_{\rm M}(\varepsilon)R_{\rm S}(\varepsilon)]+{\cal F}_{\rm free}.
\end{equation}

The integral along the real energy axis converges slowly. To convert it into a rapidly converging expression we close the integration contour in the upper half of the complex energy plane. The scattering matrices are analytic for $\operatorname{Im}\varepsilon>0$. The $\tanh$ has poles at the Matsubara frequencies $\varepsilon=i\omega_p=(2p+1)i\pi T$, with $p$-independent residue $2T$. We thus arrive at
\begin{equation}
{\cal F}=-T\sum_{p=0}^\infty\ln\det[1-R_{\text{M}}(i\omega_p)R_{\text{S}}(i\omega_p)]+{\cal F}_{\rm free},
\end{equation}
which is the result \eqref{calFgeneral} from the main text.

\end{document}